\documentclass[fleqn,usenatbib]{mnras}

\usepackage{newtxtext,newtxmath}

\usepackage[T1]{fontenc}

\DeclareRobustCommand{\VAN}[3]{#2}
\let\VANthebibliography\thebibliography
\def\thebibliography{\DeclareRobustCommand{\VAN}[3]{##3}\VANthebibliography}

\DeclareMathOperator{\sech}{sech}

\defcitealias{Fujii2018}{F18}
\defcitealias{BlandHawthorn2023}{BH23}

\usepackage{graphicx}	
\usepackage{amsmath}	

\title[Bar formation timescales]{Formation timescales for stellar bars in diverse galactic discs}

\author[M. Frosst et al.]{
Matthew Frosst$^{1}$\thanks{E-mail: matt.frosst@icrar.org}, Danail Obreschkow$^{1}$, Aaron Ludlow$^{1}$
\\
$^{1}$International Centre for Radio Astronomy Research (ICRAR), University of Western Australia, Crawley, WA 6009, Australia}

\date{Accepted XXX. Received YYY; in original form ZZZ}

\pubyear{2026}

\begin{document}
\label{firstpage}
\pagerange{\pageref{firstpage}--\pageref{lastpage}}
\maketitle

\begin{abstract}
We study the formation of stellar bars using 145 simulations of disc galaxies embedded in live and static dark matter haloes. We use the exponential bar growth timescale, $\tau_{\rm bar}$, to quantify how disc structure and kinematics regulate the onset and rate of secular bar formation. We extend previous work to thicker and more turbulent discs, motivated by those observed at high redshift ($z>1$). By revisiting several commonly used disc stability criteria -- the Efstathiou-Lake-Negroponte parameter ($\epsilon_{\rm ELN}$), the Ostriker-Peebles ratio ($t_{\rm OP}$), and the disc stellar mass fraction within 2.2 disc scale radii ($f_{\rm disc}$) -- we find that $\tau_{\rm bar}$, when expressed in terms of the disc's orbital period, follows a tight power law with each criteria. In Milky Way-like discs embedded in live haloes, bars form within a Hubble time if $f_{\rm disc} \geq 0.18$, $t_{\rm OP} \geq 0.27$, and $\epsilon_{\rm ELN} \leq 1.44$. We show discs with higher velocity dispersion experience delayed bar growth and introduce an empirical relation that correctly describes the bar formation timescales of all our live halo models. Bars in static haloes grow at roughly half the rate of those in live haloes and require substantially greater disc instability to do so.
\end{abstract}

\begin{keywords}
instabilities -- galaxies: bar -- galaxies: kinematics and dynamics -- galaxies: haloes
\end{keywords}

\section{Introduction}
Stellar bars are observed in roughly two-thirds of low redshift disc galaxies \citep[e.g.][]{Eskridge2000,Sheth2008,Masters2011}. Their prevalence has been attributed to the rapid onset of gravitational instabilities \citep{Hohl1971,OP1973}, which readily develop in kinematically cool, stellar-dominated discs (e.g., \citealp{Fujii2018, BlandHawthorn2023}, hereafter \citetalias{Fujii2018} and \citetalias{BlandHawthorn2023}, respectively). However, recent detections of barred galaxies at $z\gtrsim 2$ \citep[e.g.][]{Guo2023, LeConte2024, EspejoSalcedo2025} challenge this picture, and suggest that stellar bars can also form in turbulent discs that are more common in the early Universe \citep[][]{ForsterSchreiber2009,Stott2016, HamiltonCampos2023, Birkin2024}. How dynamically turbulent conditions influence the onset and growth of stellar bars is not well understood.

Early simulations established that isolated, self-gravitating discs are prone to bar formation \citep[e.g.][]{OP1973, ELN1982, Christodoulou1995}, leading to the development of simple analytic conditions linking galaxy properties to the onset of bar instabilities. These criteria were developed for two-dimensional discs embedded in rigid dark matter (DM) haloes, neglecting the impacts of vertical disc structure and angular momentum exchange between the disc and halo \citep[e.g.][]{Athanassoula2003, Dubinski2009, Bournaud2011, Saha2013, Collier2018, Collier2021}. Consequently, their applicability to realistic galaxies is limited \citep{Athanassoula2008, Yurin2015, Mayer2004, Sellwood2016, Izquierdo2022, Romeo2023}. 

The growth rate of stellar bars can also be used to characterise disc stability; pragmatically, discs may be considered stable if they cannot form stellar bars within a Hubble time. Using high-resolution simulations of isolated discs in live DM haloes, \citetalias{Fujii2018} showed that the formation timescale of stellar bars correlates strongly with the stellar-to-total mass fraction of the disc, $f_{\rm disc}$, measured within $2.2\, R_{\rm d}$ \citep[see also][]{Combes1981, Athanassoula1986, Carlberg1985, Valencia2017}. They found that bars form more rapidly in galaxies with higher $f_{\rm disc}$, and that bar growth within a Hubble time typically requires $f_{\rm disc} \gtrsim 0.3$. \citetalias{BlandHawthorn2023} extended this work and showed that the bar growth timescale also depends on halo mass and disc gas fraction \citep[see also][]{Verwilghen2024}. However, neither study examined the role of other key disc properties, such as disc thickness or stellar velocity dispersion, on the onset and growth rate of bars.

The kinematic and structural properties of stellar discs play key roles in bar formation \citep{Athanassoula2002,Saha2013,Athanassoula2013b,Long2014,Collier2018,Katarina2019,Ghosh2022}. In particular, both high in-plane radial velocity dispersions \citep[][]{Toomre1964, Hohl1971, Kalnajs1972, Toomre1977, Sellwood1984, Athanassoula2003, BandT2008}, and high vertical velocity dispersions in thick discs can delay or suppress bar growth \citep{Klypin2009, Aumer2017, Ghosh2022}. These trends are found in both idealised simulations of isolated discs and cosmological galaxy formation models \citep[e.g.][]{Lopez2024, Ansar2025, Fragkoudi2025, RosasGuevara2025, Frosst2025b}, and are consistent with observations showing that bars are most common in massive, dynamically cold galaxies \citep[e.g.][]{Sheth2012}. However, these effects are often neglected in analytic criteria for bar instabilities. 

Motivated by this, we analyse bar formation in $145$ high-resolution, self-consistent simulations of isolated disc galaxies. Our models systematically vary the disc-to-halo mass ratio, the ratio of the disc and halo scale radii, the disc thickness, and the in-plane velocity dispersion, and include both live and rigid DM haloes. We use these simulations to quantify how disc and halo structure influence whether a bar forms, and the timescale on which it grows. By characterising the impact of vertical and radial velocity dispersions across a broad parameter range, we extend previous results to regimes that better reflect observed disc galaxy populations. In these respects, our study complements recent work by \citet{Chen2025}, who reach similar conclusions to ours. 

This paper is organised as follows. In Section~\ref{sec:methods} we describe the simulations and analysis methods. In Section~\ref{sec:ini_results} we present our main results and compare them against global stability estimators. In Section~\ref{sec:discussion} we discuss the implications for bar formation in diverse disc galaxies across cosmic time. Our conclusions are summarised in Section~\ref{sec:conclusions}.

\section{Methods}\label{sec:methods}
In this section we present the methods used to construct initial conditions (ICs), run the simulations, and analyse bar properties.  We adopt a cylindrical coordinate system centred on the disc and align the $z$-axis with the disc's angular momentum vector. In this coordinate system, $r = (R^2+z^2)^{1/2}$ is the 3D radial coordinate, where $z$ is the height above the disc midplane, and $R$ is the distance from the $z$-axis. 

\subsection{Initial conditions}\label{ss:ics}
\begin{figure}
    \includegraphics[width=\columnwidth]{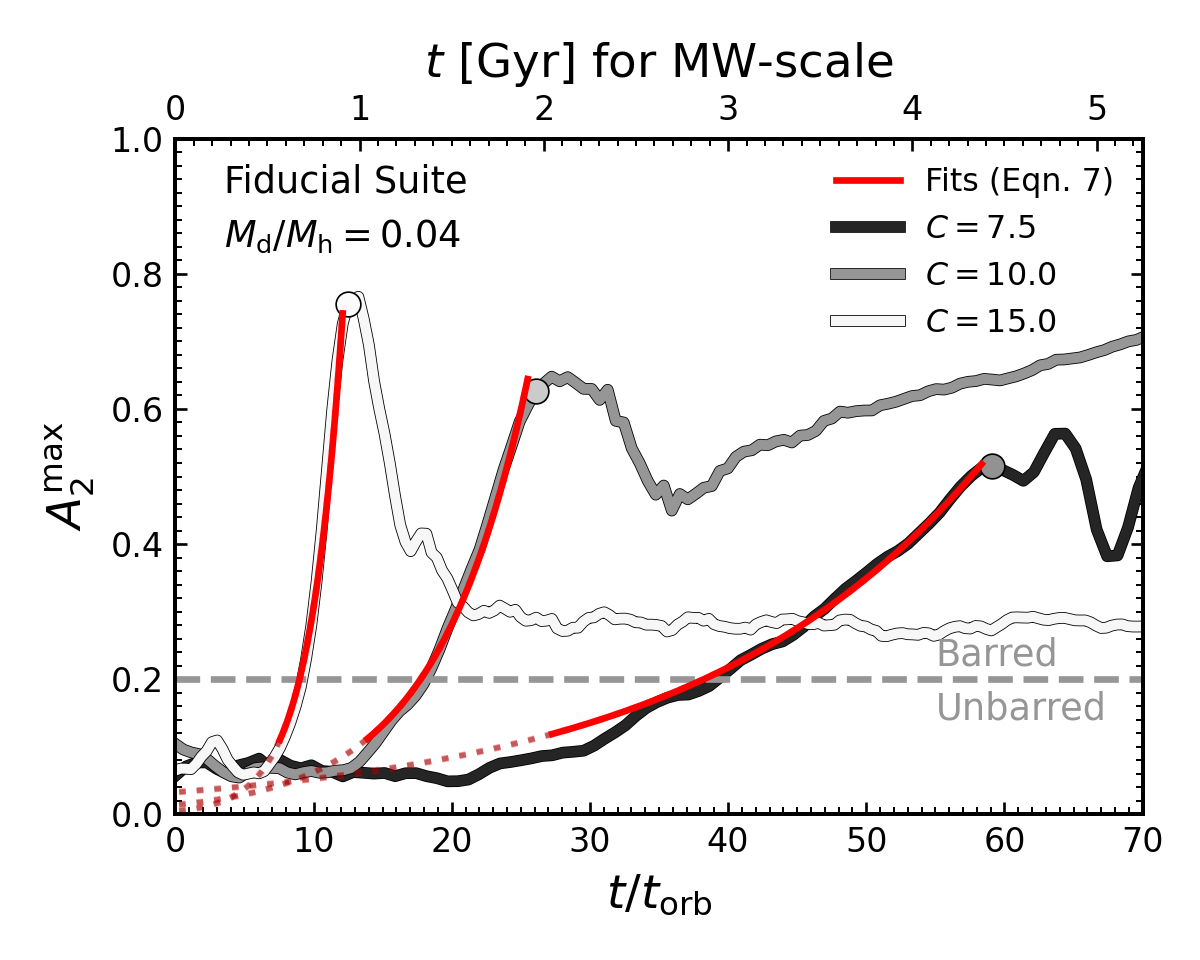}
    \caption{
        The bar strength, $A_{2}^{\rm max}$, is show as a function of time for three example models in the fiducial suite ($Q_{\rm min}=1.5$, $h_{z}=0.1$) with varied $C$. Darker lines indicate discs in more highly concentrated DM haloes, and shorter orbital times. Circles indicate the time at which the bar begins to buckle, and the assembly phase ends. The solid red lines indicate an exponential fit of Eqn.~\ref{eq:taubar} to the assembly phase for each model, the dotted red lines indicate the extrapolated fit to early times. The horizontal grey dashed line displays the delineation between barred and unbarred galaxies. }
    \label{fig:taufits}
\end{figure}

We use the \textsc{AGAMA} library \citep{Vasiliev2019} to create equilibrium initial conditions (ICs) from action-based distribution functions (DFs) for a diverse range of disc galaxies. We build our ICs as described in \citet{Frosst2024}; in short, each IC consists of a \citet{Hernquist1990} halo and a rotationally supported stellar disc, fully described by a quasi-isothermal DF and a double power-law DF, respectively \citep[see][for details]{Vasiliev2019}. After \textsc{AGAMA} iteratively converges upon the equilibrium solution for this disc-halo pair, the initial density profile of the disc can be approximately described by
\begin{equation}
    \rho_{\rm d}(R,z) = \frac{M_{\rm d}}{4\, \pi \, z_{\rm d} \, R_{\rm d}^2}\exp\left({-\frac{R}{R_{\rm d}}}\right)\sech^2\left({\frac{z}{z_{\rm d}}}\right),
\end{equation}
where $M_{\rm d}$ is the total disc mass, and $z_{\rm d}$ and $R_{\rm d}$ are the scale height and length of the disc, respectively. Similarly, the density of the halo is well described by a \citet{Hernquist1990} profile
\begin{equation}
    \rho_{\rm h}(r) = \frac{M_{\rm h}}{2\, \pi} \frac{r_{\rm h}
    }{r(r+r_{\rm h})^3},
\end{equation}
where $r_{\rm h}$ is the halo scale radius, and $M_{\rm h}$ the total halo mass. The total mass of the system is $M_{\rm tot} = M_{\rm d} + M_{\rm h}$. 

\begin{table*}
\begin{tabular}{cccccccccccc}
\hline \hline
Model type & $N_{\rm seeds}$ & $M_{\rm d}/M_{\rm h}$ & $C=r_{\rm h}/R_{\rm d}$ & $Q_{\rm min}$ & $h_{z}=z_{\rm d}/R_{\rm d}$ & $N_{\rm d}$ & $N_{\rm h}$ \\
\hline \hline
Fiducial; live halo & 3 & 0.01, 0.02, 0.03, 0.04, 0.05 & 5, 7.5, 10, 15, 20 & 1.5 & 0.1 & $10^6$ & $10^7$ \\
Fiducial; static halo & 1 & 0.01, 0.02, 0.03, 0.04, 0.05 & 5, 7.5, 10, 15, 20 & 1.5 & 0.1 & $10^6$ & \text{Static}\\
\hline
Varied $Q_{\rm min}$ & 1 & 0.01, 0.03, 0.05 & 5, 10, 20 & 1.1, 1.5, 1.7, 2.0 & 0.1 & $10^6$ & $10^7$ \\
\hline
Varied $h_{z}$ & 1 & 0.01, 0.03, 0.05 & 5, 10, 20 & 1.5 & 0.05, 0.1, 0.2 & $10^6$ & $10^7$ \\
\hline
\end{tabular}
\caption{The main properties of discs and haloes in our suites of idealised simulations. 
The first two rows describe our ``fiducial'' runs, while the final two introduce more complicated disc structure. The first column provides the names of these suites referenced throughout the paper, while the second column lists the number of random initializations ($N_{\rm seeds}$) simulated for each model. Then, from left to right, we list the four dimensionless parameters that define the ICs ($M_{\rm d}/M_{\rm h}$, $C$, $Q_{\rm min}$, $h_{z}$). The final columns show the number of particles in the disc ($N_{\rm d}$) and halo ($N_{\rm h}$), respectively. 
}
\label{tab:runs}
\end{table*}

To produce the total 6D phase space DF, \textsc{AGAMA} requires additional information on the velocity structure of the system. For the haloes we choose an isotropic velocity distribution and no net rotation \citep[see][for a discussion of how these assumptions may affect our results]{Saha2013}. For the stellar disc, we set the radial velocity dispersion, $\sigma_{r}$, to control the \citet{Toomre1964} local stability parameter, defined as
\begin{equation}
    Q(R) = \frac{\sigma_r(R)\, \kappa(R)}{3.36\, G\, \Sigma_{\rm d}(R)},
\end{equation}
where $\kappa$ and $\Sigma_{\rm d}$ are the epicyclic frequency and surface mass density of the disc, respectively. All of our discs are marginally stable against local instabilities ($Q(R)\gtrsim 1$). For each disc, we control the normalisation and radial scale of $\sigma_r(R)$ so that the Toomre-$Q$ profile reaches its minimum near the disc scale length $R_{\rm d}$ (i.e., $\sigma_{r}(R)$ declines exponentially), consistent with the typical behaviour of $Q(R)$ profiles in a wide range of potentials \citep{Obreschkow2016}. The minimum value, $Q_{\rm min}$, is therefore determined by the radial velocity dispersion. On the other hand, the disc's vertical velocity dispersion, $\sigma_{z}$, is set indirectly via the scale height, $z_{\rm d}$, and in practice also declines exponentially with $R$.

Given these parameters, \textsc{AGAMA} constructs the phase space DF of the disc-halo pair and samples it to produce our ``live halo'' ICs. All ICs use $N_{\rm d}=10^6$ disc particles, and $N_{\rm h} = 10^7$ halo particles.\footnote{Therefore, the mass ratio of individual particles in the halo and disc, $\mu\equiv m_{\rm h}/m_{\rm d}$, ranges from $2$ to $10$ in our models, but the total number of particles is sufficient to avoid the effects of spurious collisional heating \citep[e.g.][]{Ludlow2019, Ludlow2020, Ludlow2023, Wilkinson2023}.} For all models, the gravitational softening length, $\epsilon$, is set to a fixed fraction of the disc scale length, $\epsilon/R_{\rm d}=1/20$, which also ensures that $\epsilon < z_{\rm d}$. We enforce axisymmetry in the ICs by duplicating the \textsc{AGAMA} DF with a point-symmetry about the origin, then removing half of all particles at random \citep[thus reducing spurious asymmetries in the ICs;][]{Sellwood2024}. \citet{Frosst2024} established that bar formation is converged in these models, and their results agree with tests carried out by \citet{Dubinski2009} and \citetalias{Fujii2018}. 

We create suites of ICs within which we vary the disc-to-halo mass fraction, $M_{\rm d}/M_{\rm h}$, from $0.01$ to $0.05$ ($M_{\rm d}/M_{\rm h}\in\{0.01, 0.02, 0.03, 0.04, 0.05\}$), and halo-to-disc scale length ratio (concentration), $C \equiv r_{\rm h}/R_{\rm d}$, from $5$ to $20$ ($C\in\{5, 7.5, 10, 15, 20\}$). Within each suite, we set a consistent $Q_{\rm min}$ and disc scale height-to-length ratio, $h_{z} \equiv z_{\rm d}/R_{\rm d}$. Our fiducial simulation suite consists of $25$ models spanning the full range of $M_{\rm d}/M_{\rm h}$ and $C$, all with $Q_{\rm min} = 1.5$ and $h_{z} = 0.1$, respectively. Following \citet{Frosst2024}, we assess the impact of a live DM halo on bar assembly using a separate suite of static halo models. These share the same structural properties as the fiducial simulations, but the DM particles are replaced by a static \citet{Hernquist1990} potential with identical initial parameters. A summary of our simulations is provided in Table~\ref{tab:runs}. 

Our models are strictly scale-invariant in that any uniform rescaling of all mass and/or length scales results in self-similar behaviour on rescaled timescales. We exploit this invariance by presenting results in dimensionless units, including bar formation times, which we normalise by $t_{\rm orb}=2\pi R_{\rm d} / V_{\phi}(R_{\rm d})$, the orbital period at $R_{\rm d}$, where $V_{\phi}(R_{\rm d})$ is the mean disc azimuthal velocity in a cylindrical bin centred on $R_{\rm d}$ with width $\pm0.1R_{\rm d}$; note that $t_{\rm orb}$ depends on $M_{\rm d}/M_{\rm h}$ and $C$. For an intuitive reference, we occasionally quote dimensional quantities corresponding to a Milky Way-like galaxy with $M_{\rm h} = 10^{12} M_{\odot}$, $R_{\rm d}=2.2\,{\rm kpc}$, and $V_{\phi}=180\,{\rm km~s^{-1}}$ (based on the MW measurements of \citealp{Bovy2013} and estimates from GAIA data in \citealp{Katz2018}), thus $t_{\rm orb,\rm MW} \approx 75\,{\rm Myr}$.

The range of mass models and stellar kinematics that we explore is vast, and in some instances leads \textsc{AGAMA} to produce discs that deviate from the desired exponential profiles, usually by shifting $R_{\rm d}$ (which deviates from the target value by on average $\approx10$ per cent, but more notably in disc-dominated, turbulent, or thick discs). To account for this, we recompute $R_{\rm d}$ by locating the radius that encloses $26.4$ per cent of the disc mass, consistent with the scale radius of an exponential disc. Fits to the disc density profile return $R_{\rm d}$ in good agreement with these measurements. We exclude the two most disc-dominated galaxies in the $Q_{\rm min} = 2.0$ suite from our analysis, as \textsc{AGAMA} did not produce equilibrium solutions with monotonic surface density profiles for these extreme cases, a known limitation for warm discs formed from quasi-isothermal DFs \citep{Vasiliev2019}. 

\subsection{The simulation code}\label{sec:code}
We evolve our models with the {\sc GADGET-4} \citep{Springel2021} code. Gravitational accelerations between particles are computed using the Fast Multipole Method (\textsc{FMM}) with a third-order expansion ($p = 3$) as originally outlined in \citet{Greengard1987}. We use the default integration accuracy parameter $\zeta = 0.005$ and find that halving this value does not affect our results. Snapshots are output in $50\,{\rm Myr}$ ($\approx0.66t_{\rm orb,\rm MW}$) intervals for $10\,{\rm Gyr}$ ($\approx133t_{\rm orb,\rm MW}$), resulting in $200$ snapshots per run. 

To ensure reproducibility, we evolved ICs created using three distinct ``seeds'' by resampling the DFs of the fiducial live halo simulations (see Table~\ref{tab:runs}). The evolution of disc and bar properties between different random initializations of the ICs were consistent with each other, reassuring us of the robustness of the simulation results against Poisson noise. 

\subsection{Bar analysis}\label{sec:baranalysis}
We quantify the properties of stellar bars from a Fourier decomposition of the face-on stellar surface mass density following the procedures in \citet{Dehnen2022}. The strength and phase angle of the bar are obtained from the second Fourier mode in cylindrical radial bins\footnote{Following \citet{Dehnen2022}, after determining the initial number of bins, $N_\mathrm{bin}$, we add an additional $N_\mathrm{bin} - 1$ intermediate bins between the medians of all initial bins, leaving us with $2N_\mathrm{bin}-1$ overlapping bins in total. Within each bin, we calculate the $A_{2}(R)$ and $\phi_{2}(R)$ used in this work.} containing $N_{\rm part}=10^4$ particles each, and calculated in each radial bin as
\begin{align}
    &A_{\rm 2} = |\mathcal{A}_{\rm 2}| \text{, and } \\
    &\phi_{\rm 2} = \frac{1}{2}\arg (\mathcal{A}_{\rm 2}),
\end{align}
where
\begin{equation}
    \mathcal{A_{\rm 2}} = \frac{\sum_j M_{j} e^{{\rm 2}i\theta_{j}}}{\sum_j M_{j}}
\end{equation}
is the complex amplitude of the second Fourier mode. Here, $M_{j}$ and $\theta_{j}$ are the mass and azimuthal angle in the plane of the galaxy of the $j^{\rm th}$ disc particle. For this value of $N_{\rm part}$, Poisson noise fluctuates at $\sim1$ per cent. We define the bar strength, $A_{2}^{\rm max}$, as the maximum value of $A_{\rm 2}(R)$. Following standard practice in the literature, we classify galaxies as barred when $A_{2}^{\rm max} > 0.2$ \citep[e.g.,][]{Athanassoula2002, Sellwood2016, Guo2019}.

\begin{figure*}
	\includegraphics[width=\textwidth]{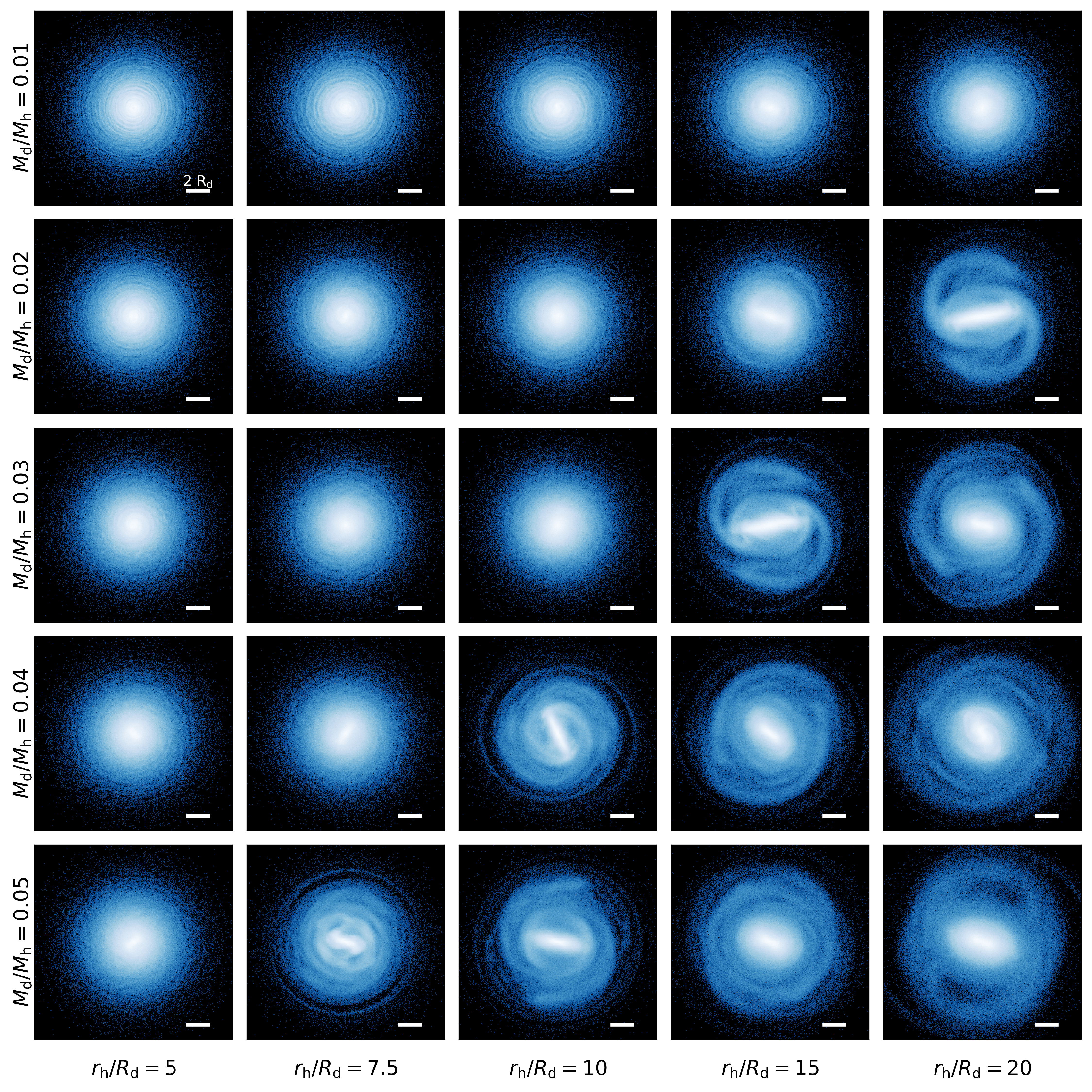}
    \caption{The face-on surface mass density projection of the stellar discs in the live halo fiducial simulation suite at $t = 2.5\,{\rm Gyr}$ for MW-scale. Each row displays a fixed disc-to-halo mass fraction, $M_{\rm d}/M_{\rm h}$, increasing from top to bottom, while each column displays a fixed halo-to-disc scale length fraction, $C$, increasing (decreasing in relative halo concentration) from left to right. All systems are run in a live Hernquist halo. The white horizontal bar indicates a scale of $2R_{\rm d}$. }
	\label{fig:grid_finalsnapshot}
\end{figure*}

Bar instabilities initially grow exponentially \citep[e.g.][]{Sellwood2014}. We therefore measure the bar growth timescale, $\tau_{\rm bar}$, by fitting an exponential to the time evolution of $A_{2}^{\rm max}$,
\begin{equation} \label{eq:taubar}
    A_{2}^{\rm max}(t) = P\times\exp\left(t / \tau_{\rm bar} \right),
\end{equation}
where $t$ is the time of the simulation snapshot, and $P$ is a free parameter to account for initial Poisson noise (see \citetalias{BlandHawthorn2023} for a similar method). This function is fit to the data only during the bar assembly phase: from the time when $A_{2}^{\rm max} \geq 0.1$ up to the first peak in the $A_{2}^{\rm max}$ evolution, which typically corresponds to the onset of buckling. In situations where the bar does not buckle before the end of the simulation, the fit is performed on all subsequent outputs. The fit is largely insensitive to the choice of fitting interval, as long as it covers the majority of the exponential growth phase.

The red lines in Fig.~\ref{fig:taufits} show example fits of Eqn.~(\ref{eq:taubar}) to the evolution of $A_{2}^{\rm max}$ for three example models in our fiducial suite; extrapolations to earlier times shown as dotted red lines. Circles indicate the time at which the bars begin to buckle, at which point the bar assembly phase ends, and we end the fit. Regardless of the disc's initial conditions, the fits are generally good, and capture the growth timescale of their bars.

\begin{figure*}
	\includegraphics[width=\textwidth]{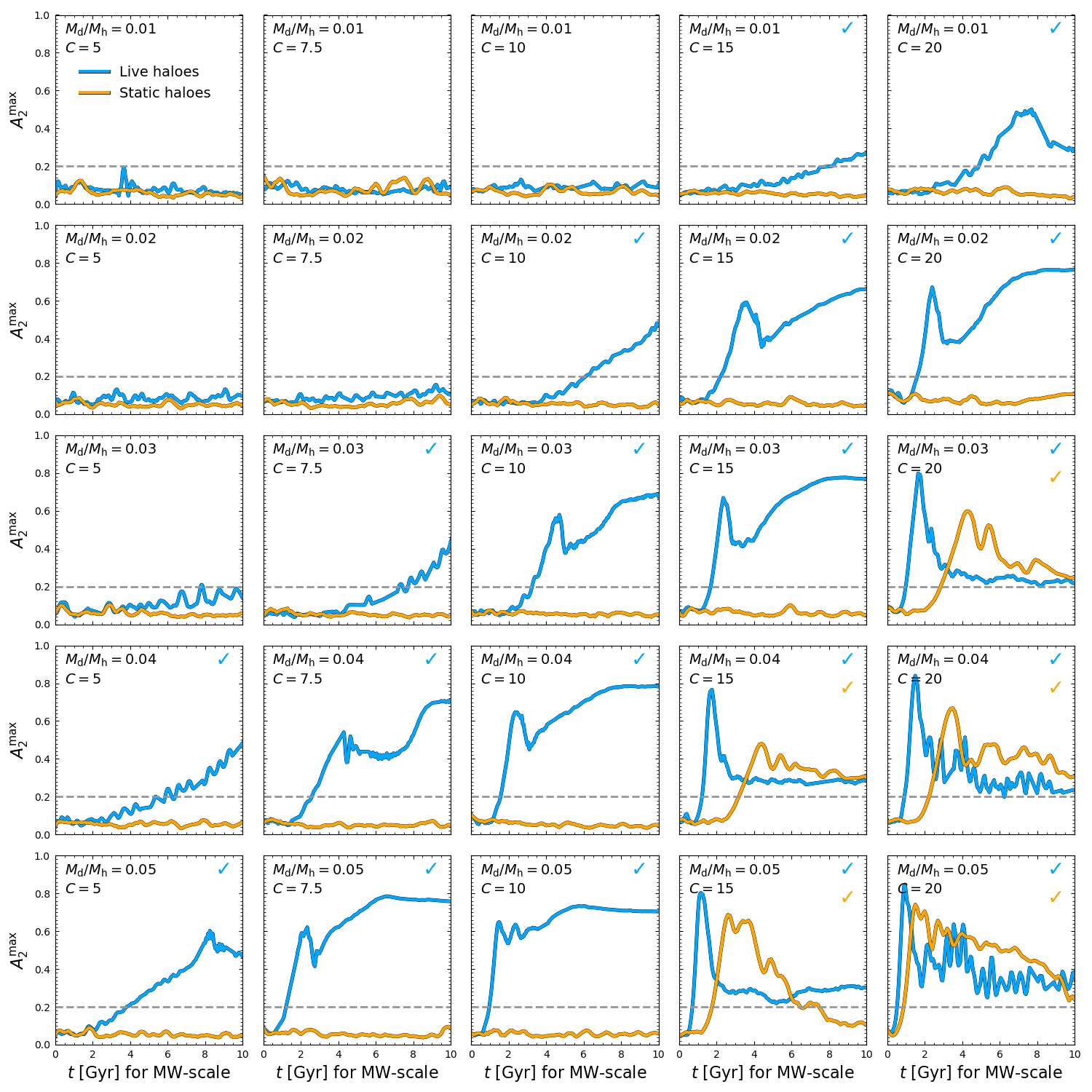}
    \caption{The bar strength, $A_{2}^{\rm max}$, as a function of time for the fiducial suite of discs in live haloes (blue) and static haloes (orange). 
    Each row displays a fixed disc-to-halo mass ratio, $M_{\rm d}/M_{\rm h}$, increasing from top to bottom, while each column displays a fixed halo-to-disc scale length ratio, $C$, increasing (decreasing in relative halo concentration) from left to right. 
    The grey dashed line indicates the $A_{2}$ amplitude above which we choose to identify bars. 
    In the top right corner of each panel we indicate with a coloured check-mark if a bar forms within the simulation runtime. 
    }
	\label{fig:grid_finalsnapshot_A2}
\end{figure*}

\begin{figure*}
    \includegraphics[width=\textwidth]{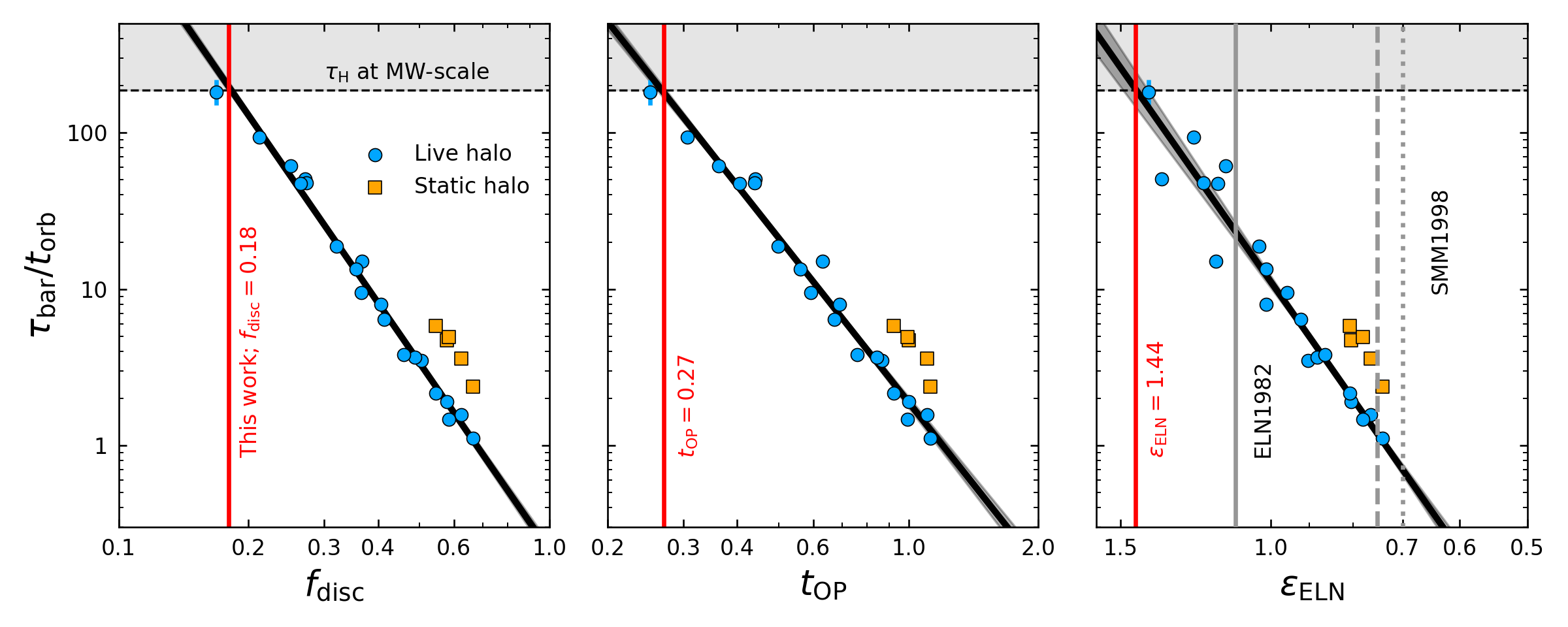}
    \caption{The normalised bar growth timescale, $\tau_{\rm bar} / t_{\rm orb}$, plotted against the central disc-to-halo mass fraction, $f_{\rm disc}$, the \citet{OP1973} criterion, $t_{\rm OP}$, and the \citet{ELN1982} criterion, $\epsilon_{\rm ELN}$, from left to right, respectively, shown in log-log space. The dashed horizontal line and grey shaded region at the top of each panel indicate timescales greater than a Hubble time, $\tau_{\rm H}$. Simulations with live haloes are plotted as blue circles, while simulations in static haloes are plotted as orange squares. Fits to the live halo data are shown as thick black lines, with dark shaded regions showing the fit errors. The criteria for disc stability defined by \citet[][labelled ELN1982]{ELN1982}, $\epsilon_{\rm ELN} = 1.1$, is shown as a grey vertical line, while those from \citet[][labelled SMM1998]{Syer1998} are plotted as dashed and dotted grey lines. The vertical, solid red lines show the stability threshold below which bars form in our fiducial suite within $\tau_{\rm H}$.}
    \label{fig:original_ELN}
\end{figure*}

\section{Results}\label{sec:ini_results}
\subsection{Overview of bar growth in fiducial simulation suite}
In Fig.~\ref{fig:grid_finalsnapshot}, we show the face-on stellar mass density distribution of our live-halo fiducial suite at $t=2.5\,{\rm Gyr}$. From left to right, panels correspond to increasing stellar mass concentration $C=r_{\rm h}/R_{\rm d}$, while from top to bottom, they correspond to increasing stellar mass fraction, $M_{\rm d}/M_{\rm h}$; the most strongly halo-dominated model is in the top left, and the most disc-dominated model is in the bottom right. After only a short period, bars have already formed in the most disc-dominated galaxies, and are also starting to form in more halo-dominated systems. Fig.~\ref{fig:grid_finalsnapshot} shows that there is a smooth progression of more rapidly forming bars moving diagonally across this parameter space, from the top left to bottom right. 

Blue lines in Fig.~\ref{fig:grid_finalsnapshot_A2} show the full evolution of $A_{2}^{\rm max}$ for these fiducial, live halo models (following the same layout as Fig.~\ref{fig:grid_finalsnapshot}). Bars form earliest and strongest in galaxies whose ICs are most disc-dominant (i.e., $A_{2}^{\rm max}$ increases most rapidly, and peaks at the highest values in galaxies with the largest $M_{\rm d}/M_{\rm h}$ and $C$), and the initial $A_{2}^{\rm max}$ growth is always exponential, with the exception of Poisson noise. Almost all models show a rise in $A_{2}^{\rm max}$ by the end of the simulation (only four models do not exhibit a statistically significant evolution in $A_{2}^{\rm max}$ beyond the Poisson noise floor), and most go on to form bars ($76$ per cent have $A_{2}^{\rm max} > 0.2$ in the last snapshot). Only the most halo-dominated discs appear to be long-term stable. In contrast, the orange lines in Fig.~\ref{fig:grid_finalsnapshot_A2} show the time evolution of $A_{2}^{\rm max}$ for the corresponding models in static haloes. In these cases, bars always form later and reach a lower maximum $A_{2}^{\rm max}$ compared to identical discs in live haloes \citep[see also][]{Sellwood2016, Sellwood2025, Frosst2024}. Ultimately, the vast majority of discs in static haloes do not form bars in the time available and experience very little evolution in $A_{2}^{\rm max}$ beyond the Poisson noise present in the initial conditions. 

\subsection{Bar growth dependence on global stability metrics}\label{sec:results_ini}
On analytical grounds, \citet{OP1973} found that the stability of a disc against bar formation depends on the ratio between rotational kinetic energy and gravitational binding energy, which, assuming virial equilibrium conditions, can be expressed as
\begin{equation}\label{eq:top}
    t_{\rm OP} = \frac{T}{|2\Pi + 2T|},
\end{equation}
where $T$ is the rotational kinetic energy and $\Pi$ the random kinetic energy. Using early $N$-body simulations (with then large, but now understood to be insufficient numbers of particles of $N=500$), they found that bars form if $t_{\rm OP}\geq0.14$. For consistency with other stability criteria, we measure $t_{\rm OP}$ within $R\leq 2.2R_{\rm d}$, adopting the practical definitions $T=\tfrac{1}{2}\langle v_\phi\rangle^2$ and $\Pi=\tfrac{1}{2}\langle v_\phi-\langle v_\phi\rangle\rangle^2+\langle v_R\rangle^2+\langle v_z\rangle^2$, where $\langle...\rangle$ are mass-weighted means.

Later, \citet{ELN1982} introduced an alternative, but related disc stability metric for thin discs in circular rotation,
\begin{equation}
 \label{eq:eln}
 \epsilon_{\rm ELN} = \frac{V_{\rm max}}{(GM_{\rm d}/R_{\rm d})^{1/2}},
\end{equation}
where $V_{\rm max}$ is the maximum of the circular velocity. This parameter describes the importance of the disc's self-gravity relative to the gravity of the entire system: models with lower $\epsilon_{\rm ELN}$ will have higher relative self-gravity, and thus a higher the chance for a bar instability to grow. Using 2D $N$-body simulations of isolated flat discs (again limited by low particle numbers of $N=20,000$), they found that bars form when $\epsilon_{\rm ELN}<1.1$. This result was also restricted by the assumption of a static halo rather than a responsive one, and an unrealistic DM density profile. Subsequently, \citet{Syer1998} found that 3D discs in static Hernquist haloes are slightly more stable, forming bars only below $\epsilon_{\rm ELN} \lesssim 0.7$, though this result again neglects the importance of co-evolving haloes. 

Interestingly, an even simpler metric than $t_{\rm OP}$ and $\epsilon_{\rm ELN}$ has recently become established as remarkably good indicator of bar formation: the disc-to-total mass ratio (\citealp{Athanassoula1986}, see also \citealp{Combes1981, Athanassoula2002}), which is often measured in a spherical aperture of radius of $2.2\, R_{\rm d}$ (\citetalias{Fujii2018} and \citetalias{BlandHawthorn2023}), i.e.\ the radius at which a self-gravitating exponential disc reaches its maximum circular velocity \citep[see also][]{Widrow2008,Devergne2020}. Here, we adopt the definition
\begin{equation} \label{eq:eta}
    f_{\rm disc} = \frac{M_{\rm disc}(r\leq 2.2R_{\rm d})}{M_{\rm tot}(r\leq2.2R_{\rm d})},
\end{equation}
noting that key proponents of this metric (\citetalias{Fujii2018} and \citetalias{BlandHawthorn2023}) express $f_{\rm disc}$ as the square of the circular velocity ratio, which is identical to Eqn.~(\ref{eq:eta}) in the Keplerian approximation for the circular velocity.

Fig.~\ref{fig:original_ELN} shows the relation between the three metrics given in Eqns.~(\ref{eq:top}--\ref{eq:eta}) and the exponential bar growth timescale $\tau_{\rm bar}$ for our fiducial live and static halo simulation suites (blue circles and orange squares, respectively). We plot the bar growth timescale (normalised by the disc orbital time, i.e. $\tau_{\rm bar}/t_{\rm orb}$) against $f_{\rm disc}$, $t_{\rm OP}$, and $\epsilon_{\rm ELN}$ from left to right, respectively. Error bars for $\tau_{\rm bar}/t_{\rm orb}$ are included for all bar forming galaxies, but are smaller than the dot sizes in most cases. The horizontal dashed line denotes the times above which the bar growth timescale exceeds a Hubble time, $\tau_{\rm H}=13.78\,{\rm Gyr}$, for MW-like galaxy and halo scales (Section~\ref{ss:ics}). The $x$-axes are oriented such that galaxies predicted to be more prone to bar formation appear further to the right in all three panels.

Fig.~\ref{fig:original_ELN} shows that the bar growth timescales are tightly correlated with all three global stability parameters, evaluated at the initial conditions. The numerical data shown here is the fiducial suite, where we vary $M_{\rm d}/M_{\rm h}$ and halo-to-disc scale length ratio (concentration), $C$, but keep the dispersion metrics ($Q_{\rm min}$ and $h_{z}$) fixed. The normalised growth times, $\tau_{\rm bar}/t_{\rm orb}$, appear to vary as power laws of $f_{\rm disc}$, $t_{\rm OP}$, and $\epsilon_{\rm ELN}$. The best-fitting power laws, in terms minimising $\chi^2$ in log space, are
\begin{align}
    &\tau_{\rm bar}/t_{\rm orb} = 0.21 \, f_{\rm disc}^{-4.00} \label{eq:eta_fit}, \\
    &\tau_{\rm bar}/t_{\rm orb} = 1.90 \, t_{\rm OP}^{-3.48}, \text{ and,} \label{eq:top_fit} \\
    &\tau_{\rm bar}/t_{\rm orb} = 11.22\, \epsilon_{\rm ELN}^{7.77}. \label{eq:eln_fit}
\end{align}
These power laws, shown as thick black lines in Fig.~\ref{fig:original_ELN}, are remarkably tight, underscoring the dominant role of the disc's self-gravity in regulating the secular assembly of bars. Small changes to the radius at which these stability parameters are measured will change the values obtained by these fits, however, we find $2.2R_{\rm d}$ minimises the scatter along these relations. We do not perform equivalent fits for the static halo models, as $\tau_{\rm bar}$ can be determined accurately in only a few cases. Nonetheless, our static halo models consistently form bars more slowly than their live halo counterparts by about a factor of 2, and require stronger initial disc instabilities to do so, highlighting the stabilising influence of rigid DM haloes.

Eqns.~(\ref{eq:eta_fit}--\ref{eq:eln_fit}) can be used to estimate threshold values for bars to form within a Hubble time, $\tau_{\rm H}$, provided a choice of physical dimensions setting $t_{\rm orb}$. For example, for MW-scale systems (see Section~\ref{ss:ics}), the fit of Eqn.~(\ref{eq:eta_fit}) predicts that bars form within $\tau_{\rm H}$ if $f_{\rm disc} \gtrsim 0.18$, slightly lower than the limit of $\gtrsim 0.3$ found in \citetalias{Fujii2018} and \citetalias{BlandHawthorn2023} \citep[but see also][]{Widrow2008, Sellwood2014, Valencia2017, Devergne2020}. Similarly, our fits suggest that bars can form within $\tau_{\rm H}$ when $t_{\rm OP} > 0.27$. This is slightly higher than the threshold of $t_{\rm OP} > 0.14$ found by \citet{OP1973} beyond which discs can remain indefinitely stable to bar formation. Likewise, our fits suggest that a disc must have $\epsilon_{\rm ELN} < 1.44$ for a bar to form within $\tau_{\rm H}$, substantially higher than the thresholds of $\epsilon_{\rm ELN} = 1.1$ or $0.7$ proposed by \citet{ELN1982} and \citet{Syer1998}, respectively. We therefore produce many clear counter examples that violate previous stability criteria, if given sufficient time to evolve in isolation. 

Of the three considered metrics, $f_{\rm disc}$ provides the tightest fit to $\tau_{\rm bar}/t_{\rm orb}$ and hence appears to be a dominant driver of secular bar formation. The relationship between $\tau_{\rm bar}$ and $f_{\rm disc}$, first established by \citetalias{Fujii2018}, has come to be known as the ``Fujii relation'' (see \citetalias[][]{BlandHawthorn2023}).

\subsection{Comparison to previous studies}\label{ss:comp}
\begin{figure}
    \includegraphics[width=\columnwidth]{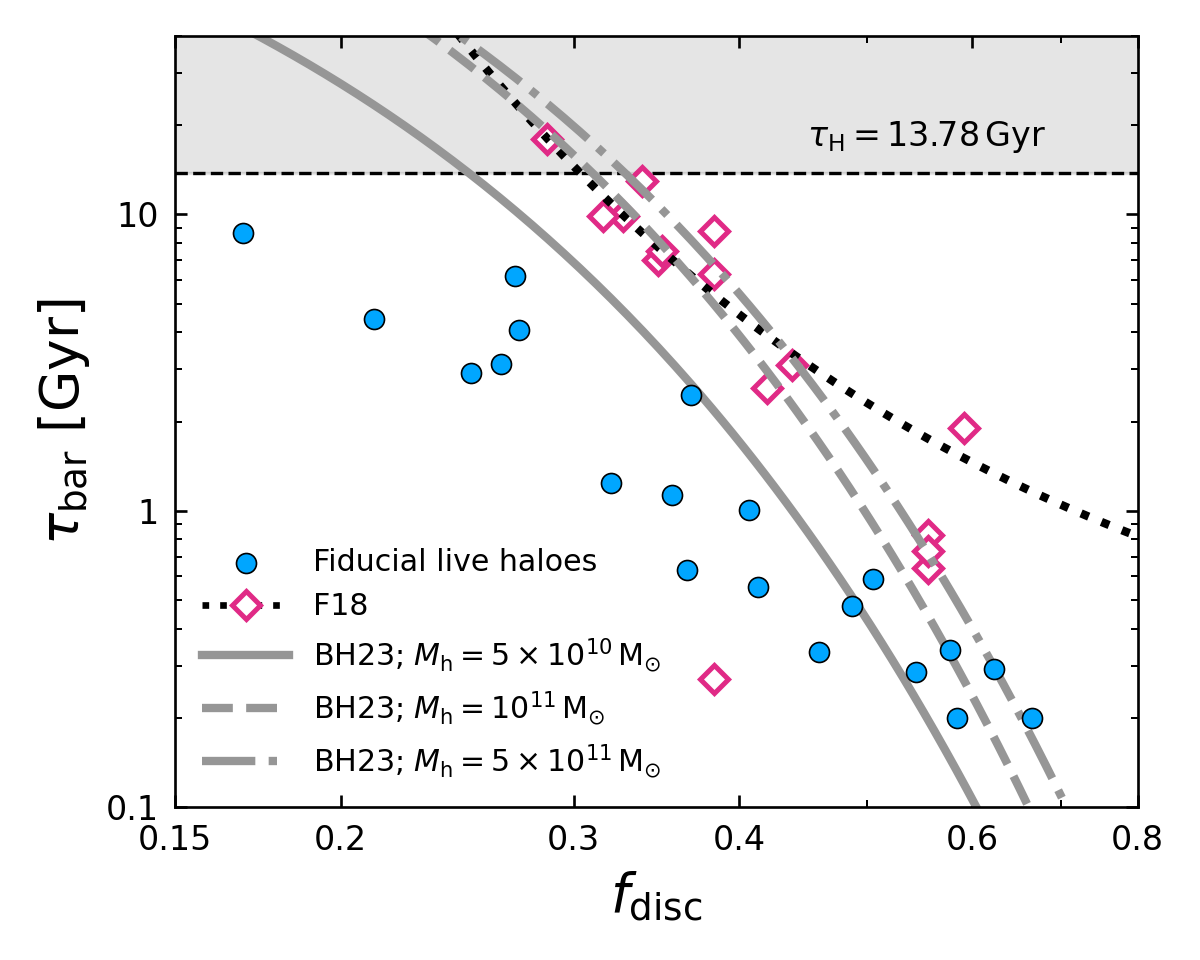}
    \caption{Overview of the bar growth timescale $\tau_{\rm bar}$ as a function of $f_{\rm disc}$, for the fiducial live halo runs of this work, the runs of \citetalias{Fujii2018} and those of \citetalias{BlandHawthorn2023}. The data of \citetalias{BlandHawthorn2023} contains three subsets of different halo masses, which we show using their their three corresponding fits (grey lines). The data from \citetalias{Fujii2018} are shown as pink diamonds, and a similar fit to this data is shown as the black dotted line. This figure does not represent a like-for-like comparison, but illustrates the differences caused by various physical, numerical and post-processing choices (see Section~\ref{ss:comp}).
    }
    \label{fig:comparison}
\end{figure}

Fig.~\ref{fig:comparison} compares our results with those of \citetalias{Fujii2018} and \citetalias[][]{BlandHawthorn2023}. To remain consistent with the dimensional timescales adopted in those studies, we present our results in terms of the absolute bar growth time, $\tau_{\rm bar}$, rather than the dimensionless timescale $\tau_{\rm bar}/t_{\rm orb}$ used in the remainder of this work. As a result, $\tau_{\rm bar}$ depends on the (necessarily somewhat arbitrary) choice of physical scales used in our simulations. Specifically, we set the halo mass and disc scale to $M_{\rm h}=10^{12}\,M_\odot$ and $R_{\rm d}=2\,\mathrm{kpc}$, while varying the disc-to-halo mass ratio $M_{\rm d}/M_{\rm h}=0.01\!-\!0.05$ and the concentration parameter $C=r_{\rm h}/R_{\rm d}=5\!-\!20$ (Table~\ref{tab:runs}). For this set of fiducial models with live haloes, the corresponding orbital time spans $\sim 50$--$180\,\mathrm{Myr}$. As a consequence of this scaling, these comparisons are illustrative rather than predictive. 

There are several possible reasons why our results in Fig.~\ref{fig:comparison} differ from previous findings. First, $\tau_{\rm bar}$ naturally scales with the overall physical dimensions of the system; hence, not normalising by $t_{\rm orb}$ makes the the relationship between bar formation time and $f_{\rm disc}$ scale dependent. This is demonstrated directly by the spread between the three fits of \citetalias[][]{BlandHawthorn2023} and can also be seen in the increased scatter of our data points (blue circles) compared to the tight power law relation of the same data in normalised units (Fig.~\ref{fig:original_ELN}, left). Moreover, without normalising by $t_{\rm orb}$, the slope of the $f_{\rm disc}$--$\tau_{\rm bar}$ relation depends on what physical quantities are held fixed while varying $f_{\rm disc}$ ($R_{\rm d}$ in our case, $r_{\rm h}$ in \citetalias[][]{BlandHawthorn2023}, and $M_{\rm h}$ in both).

\begin{figure*}
    \includegraphics[width=\textwidth]{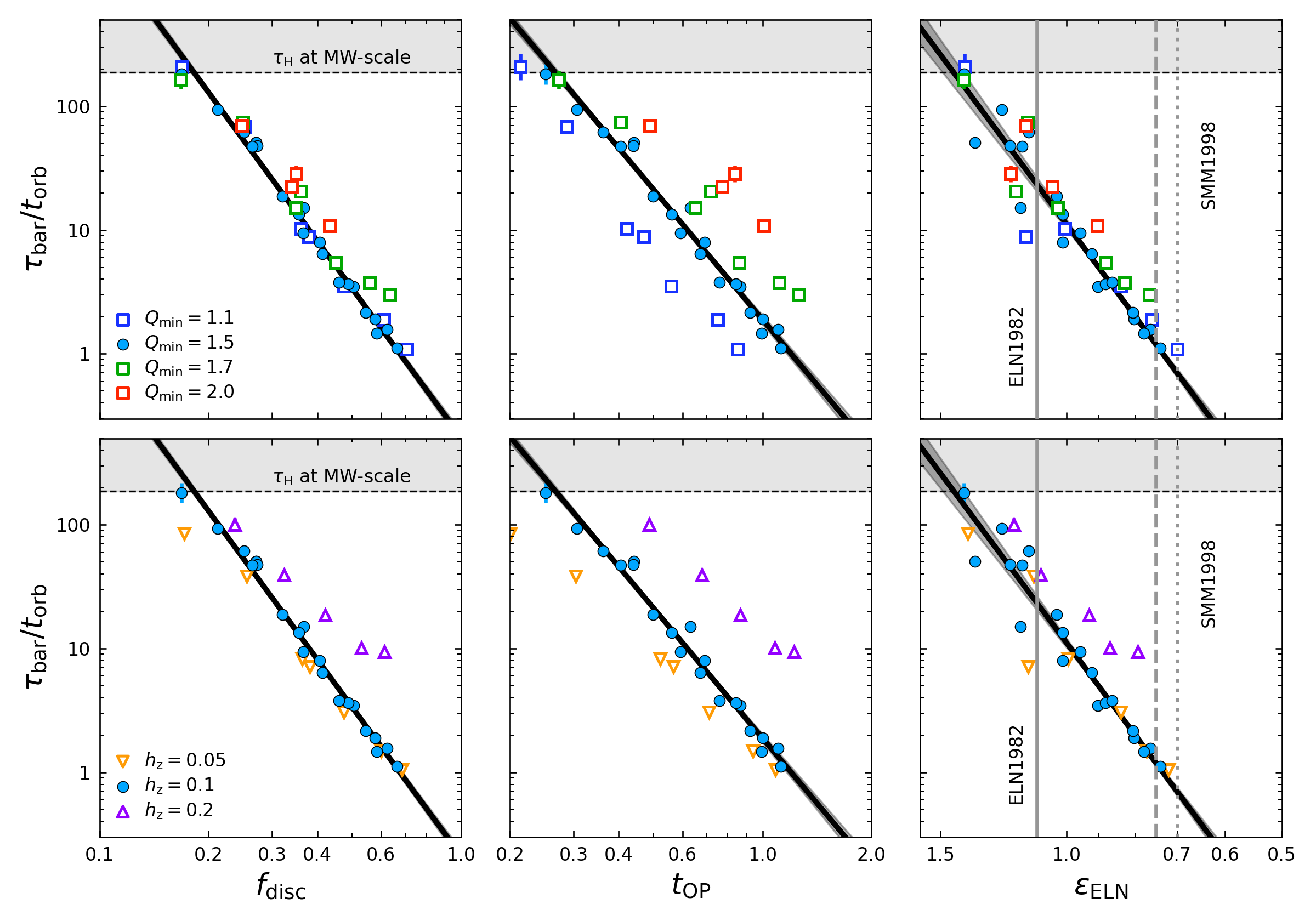}
    \caption{Same as Fig.~\ref{fig:original_ELN}, but for variations in the minimum Toomre $Q$ value, $Q_{\rm min}$, and scale height, $h_{z}$, in the top and bottom panels, respectively. The fiducial suite ($Q_{\rm min} = 1.5$ and $h_{z}=0.1$) are again shown as blue circles, fit with the thick black line. Fit errors are shown as a dark shaded region. In the top row, from lower to higher $Q_{\rm min}$, more turbulent discs ($Q_{\rm min}=2.0$) are plotted as red squares, intermediately turbulent discs ($Q_{\rm min}=1.7$) as green squares, and the least turbulent discs are plotted as blue squares ($Q_{\rm min}=1.1$). Similarly, in the bottom row, the thin discs ($h_{z} = 0.05$) are plotted as orange downward facing triangles, and the thick discs ($h_{z}=0.2$) are plotted as purple upward facing triangles.
    }
    \label{fig:QH_stability}
\end{figure*}

Second, \citetalias{Fujii2018} do not use an exponential bar formation time, $\tau_{\rm bar}$, but instead adopt the time $\tau_{0.2}$ at which $A_2^{\rm max}$ first reaches a threshold value of 0.2. Unlike the exponential time, this latter choice depends strongly on numerical resolution, which determines the initial Poisson noise from which bar modes grow: the larger the number of disc particles, $N_{\rm d}$, the smaller the initial $A_2$ noise and hence the longer the time to reach $A_2=0.2$ \citep[e.g.][]{Dubinski2009}. Importantly, this statement applies even for very large numbers of particles ($N_{\rm d}=10^6$ in our case, $N_{\rm d}=8.3\times10^6$ in \citetalias{Fujii2018}), where other bar properties are converged \citep{Frosst2024}. Comparing $\tau_{\rm bar}$ to $\tau_{0.2}$ in our simulations and extrapolating to the higher resolution of \citetalias{Fujii2018}, we estimate that the difference in defining the bar formation time explains most of the offset between our data (blue cirles) and those of \citetalias{Fujii2018} (pink diamonds).

Third, both \citetalias{Fujii2018} and \citetalias[][]{BlandHawthorn2023} include bulges in their models. This extra central mass is expected to add a stabilising spherical potential, which may slightly delay bar formation \citep[e.g.][]{Saha2013, Kataria2018}. Similarly, our use of a \citet{Hernquist1990} halo also differs from \citetalias{Fujii2018} and \citetalias{BlandHawthorn2023}, who use Navarro-Frenk-White \citep[NFW;][]{NFW1997} profiles. The two profiles are nearly identical over the radial extend of the disc but diverge at large radii, where $\rho_{\rm h}\propto r^{-4}$ for Hernquist haloes and $\rho_{\rm h}\propto r^{-3}$ for NFW haloes. However, because bars exchange angular momentum with the DM halo primarily near the co-rotation radius where the haloes differ negligibly \citep{Athanassoula2003}, we expect the choice of halo profile to have little impact on $\tau_{\rm bar}$. 

There are also other reasons that may affect the comparison between the different samples in Fig.~\ref{fig:comparison}, such as differences in halo kinematics and disc dispersion. For instance, in Sections~\ref{sec:veldisp} we will show how even small deviations in the disc structure and kinematics can have a significant impact on $\tau_{\rm bar}$ at fixed $f_{\rm disc}$.

\subsection{The effect of velocity dispersion on bar growth} \label{sec:veldisp}
Velocity dispersion is expected to play a critical role in regulating bar formation. The radial velocity dispersion, imprinted in the Toomre stability parameter, $Q$, can reduce the responsiveness of the disc to non-axisymmetric perturbations \citep[e.g.][]{Toomre1964, Kalnajs1972, Athanassoula2003, BandT2008}. Similarly, the vertical velocity dispersion, which determines the disc scale height, $z_{d}$, provides vertical support against self-gravity and drives stellar orbits out of the disc plane \citep[e.g.][]{Klypin2009, Ghosh2022} in a mechanism not captured by classical two-dimensional stability criteria. In both cases, higher velocity dispersions are expected to inhibit bar formation \citep[e.g.][]{Athanassoula2003}. 

\begin{figure*}
    \includegraphics[width=\textwidth]{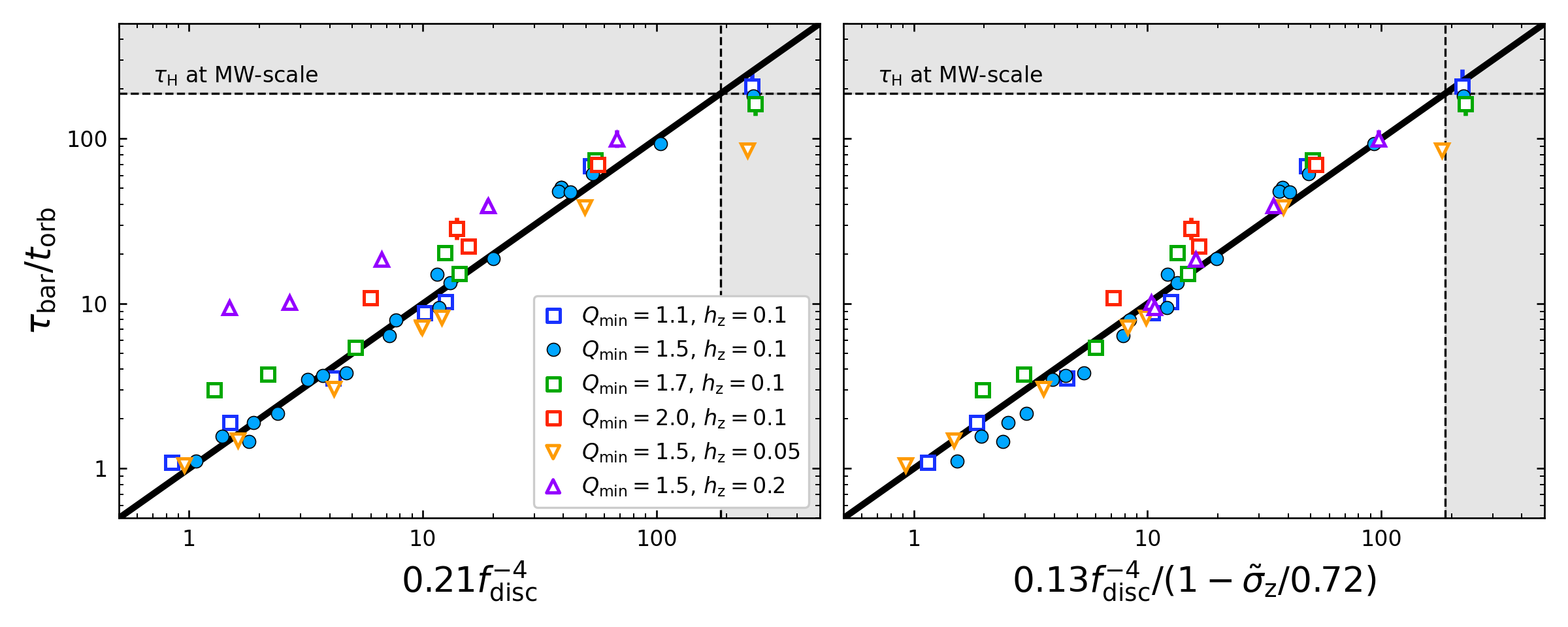}
    \caption{The normalised bar growth timescale, $\tau_{\rm bar}/t_{\rm orb}$ plotted against Eqn.~\ref{eq:eta_fit} (left) and Eqn.~\ref{eq:correction} (right) shown in log-log space. Points are coloured as in Fig.~\ref{fig:QH_stability}. Dashed lines and grey shaded regions indicate timescales greater than a Hubble time, $\tau_{\rm H}$. Solid black lines show the 1:1 relationships between the x- and y-axes.
    }
    \label{fig:corrections}
\end{figure*}

To quantify the effect of dispersion, we introduce five new suites of isolated disc galaxy models that, when used in conjunction with the fiducial models, systematically vary $Q_{\rm min}\in\{1.1, 1.5, 1.7, 2.0\}$ and $h_{z}\in\{0.05, 0.1, 0.2\}$. Each new suite spans a range of disc-to-halo mass ratios ($M_{\rm d}/M_{\rm h}\in\{0.01, 0.03, 0.05\}$) and halo concentrations ($C\in\{5, 10, 20\}$), and all use live DM haloes. The time evolution of the bar strengths of these new models are displayed in Fig.~\ref{fig:varyQall} and Fig.~\ref{fig:varyHzall}, respectively, and their structural parameters and simulation properties are available in Table~\ref{tab:runs}. These models were generated and run with the same techniques as the fiducial simulation suite discussed in Sec.~\ref{sec:results_ini}. Together, these simulations allow us to assess how velocity dispersion influences the bar growth timescale, $\tau_{\rm bar}$, across $95$ distinct galaxy models.

Fig.~\ref{fig:QH_stability} shows $\tau_{\rm bar} / t_{\rm orb}$ as a function of $f_{\rm disc}$, $t_{\rm OP}$, and $\epsilon_{\rm ELN}$, for multiple suites where $Q_{\rm min}$ (top panel) and $h_{z}$ (bottom panel) are varied. The different samples also exhibit a power law relations between $\tau_{\rm bar}/t_{\rm orb}$ and the disc stability parameters, with similar slopes but different normalisations than the fiducial runs.

As expected, increased velocity dispersion generally leads to higher normalisations and slower bar formation. For instance, galaxies with $f_{\rm disc}\approx0.4$ form bars about $2.7$-times faster when $Q_{\rm min} = 1.1$ than when $Q_{\rm min} = 2.0$ (for $h_{z} = 0.1$), and about $3.1$-times faster when $h_{z}=0.05$ than when $h_{z}=0.2$ (for $Q_{\rm min} = 1.5$). These differences correspond to changes in $\tau_{\rm bar}$ of $\sim\!1-2\,{\rm Gyr}$ in MW-like discs, highlighting the high sensitivity of bar formation to relatively small variations in disc kinematics.

The marked slowdown in bar growth for high $Q_{\rm min}$ is in line with analytic expectations and previous simulation results \citep[e.g.][]{Toomre1977, Sellwood1984, Athanassoula1986, Athanassoula2003}, but our results extend these findings to more diverse and robust live halo models. Recent cosmological simulations have likewise shown that barred galaxies have lower values of $Q$ relative to their unbarred counterparts \citep{Fragkoudi2025, RosasGuevara2025}. Our results also align with studies of bar formation in thick discs, which find that higher vertical velocity dispersions can inhibit or delay bar formation \citep[see, e.g.,][]{Klypin2009, Aumer2017, Ghosh2022}, which has also been tentatively confirmed by cosmological simulations \citep[e.g.][]{Izquierdo2022, Frosst2025b}. 

These results have important implications for interpreting bar formation at high redshift. Observations indicate that galactic discs become progressively thicker and more vertically turbulent towards earlier cosmic times, with elevated $\sigma_z$ and radial dispersions \citep[e.g.][]{Elmegreen2006, Jimenez2023}, conditions that our models show can substantially extend bar growth timescales and, in some cases, suppress secular bar formation altogether. However, our results also show that discs with relatively high radial dispersions ($Q_{\rm min}=2.0$) can still form bars within a few Gyr provided they are sufficiently disc-dominated, suggesting that bar formation remains viable in DM-poor, turbulent high-redshift galaxies. Moreover, high-redshift galaxies are subject to more frequent tidal perturbations, which can trigger bars even in otherwise stable systems \citep{Frosst2025b}. 

Finally, as for the fiducial models, bars never form faster than $t_{\rm orb}$, indicating a lower limit to the bar formation timescale that persists regardless of the initial properties of discs. This limit is approximately consistent with \citetalias{BlandHawthorn2023} \citep[and is expected from linear theory, e.g.,][]{Hamilton2025}.

\subsection{Modelling dispersion effects on the bar formation timescale}
We have shown that in our collisionless simulations the central stellar-to-total mass ratio, $f_{\rm disc}$, is a robust predictor of the normalised bar formation timescale, $\tau_{\rm bar}/t_{\rm orb}$. The two dimensionless quantities follow a power law relation given in Eqn.~(\ref{eq:eta_fit}) and shown in the left panels of Figures~\ref{fig:original_ELN} and \ref{fig:QH_stability}.

For the following discussion, it is convenient to visualise this power law between $f_{\rm disc}$ and $\tau_{\rm bar}/t_{\rm orb}$ as a one-to-one relation between the left and right side of Eqn.~(\ref{eq:eta_fit}), as in Fig.~\ref{fig:corrections} (left panel). In doing so, the $x$-axis shows the predicted normalised timescale, and the $y$-axis the observed one. As expected from the discussion of Section~\ref{sec:veldisp}, the fiducial models (blue filled circles) lie closely on this relation. However, deviations are seen if the velocity dispersion of the disc is varied, which we achieved by varying $Q_{\rm min}$ and $h_{z}$ in the \textsc{AGAMA} ICs (correspondingly $\sigma_{R}$ and $\sigma_{z}$, see Section~\ref{ss:ics}).

By looking closely at the correlations between $\tau_{\rm bar}/t_{\rm orb}$ and various dispersion metrics, we discovered that all our simulations, including those with varying velocity dispersion, are well described by the extended fitting formula,
\begin{equation} \label{eq:correction}
    \frac{\tau_{\rm bar}}{t_{\rm orb}} = 0.13\, f_{\rm disc}^{-4}\left(\frac{1}{1 - \tilde{\sigma}_{z}/0.72}\right),
\end{equation}
where $\tilde{\sigma}_{z} = \sigma_{z}/V_{\phi}$ is the ratio of the vertical velocity dispersion to azimuthal velocity measured at $R_{\rm d}$ (in a cylindrical bin of width $\pm0.1R_{\rm d}$, though varying the bin size by a factor of $\pm2$ does not change the results). This relation is visualised in the right panel of Fig.~\ref{fig:corrections}.

The inclusion of the empirical $\tilde{\sigma}_{z}$ term captures how thicker discs suppress the growth of non-axisymmetric instabilities for the physical reasons discussed in Section~\ref{sec:veldisp}. Interestingly, the same term also improves the fits for the models where $Q_{\rm min}$ is varied, keeping $h_z$ fixed. This may have to do with the fact that velocity dispersions in different dimensions are normally correlated, and thus forcing an increase of the velocity dispersion in the radial direction (by increasing $Q_{\rm min}$ in \textsc{AGAMA}) will leak some random kinetic energy into the vertical direction captured by $\tilde{\sigma}_{z}$. Overall, $\tilde{\sigma}_{z}$ can be interpreted as a dimensionless metric of how ``kinematically hot'' the disc is: larger values occur when stars spend less time on circular orbits near the mid-plane.

Eqn.~\ref{eq:correction} indicates that bar formation is suppressed for $\sigma_z \gtrsim 0.72$. Such large vertical dispersions correspond to thick discs that are unlikely to host stellar bars in the absence of external perturbations \citep{Frosst2025b}. The $\sigma_z$ term in Eqn.~\ref{eq:correction} approximately captures the extended bar formation timescales associated with strong random vertical stellar motions, which act to smooth non-axisymmetric modes before they can grow. We emphasize that this is an empirical result derived from our analysis. Other parameters, such as the disc scale height $h_z$, also yield improved fits to the data, but $\sigma_z$ provides the best overall performance among the variables we tested. 

\section{Discussion} \label{sec:discussion}
\subsection{Caveats to our analysis}
There are several caveats to this work that must be acknowledged. Our simulations neglect star formation, energetic feedback, and galactic components such as bulges, stellar haloes, DM substructure, and gas discs. This deliberate simplicity allows us to better understand the dynamical processes relevant for bar formation, but limits the realism of our models. In particular, disc galaxies, especially those at higher redshifts, are anticipated to contain substantial gas reservoirs, often exceeding $\approx 50$ per cent of the baryonic mass \citep[e.g.][]{Daddi2010, Tacconi2010}. Previous studies have shown that gas discs can both inhibit and accelerate bar formation depending on its distribution and mass relative to the stellar disc. For example, \citet[][]{Bournaud2005} and \citet{Wozniak2009} found that high gas fractions suppress bar growth, whereas \citetalias{BlandHawthorn2023} and \citet{Verwilghen2024} showed that moderate gas fractions ($\approx10$--$20$ per cent of the stellar disc mass) can halve the bar growth timescale relative to the \citetalias{Fujii2018} relation \citep[see also][]{Berentzen1998, Athanassoula2013b, BlandHawthorn2024}. Other works report either accelerated \citep{Robichaud2017} or negligible \citep{Berentzen2007} effects. This discrepancy may be related to the feedback prescriptions (or lack thereof) of these simulations, which can induce turbulence into the gas disc, accelerating the growth of bar-forming perturbations \citep[e.g.][]{Robichaud2017, BlandHawthorn2024}. Another possible reason for the discrepancy between different conclusions is that if working in dimensional units, systems with higher orbital times (at $R_{\rm d}$) generally form their bars more slowly. Following the results of this work it is imperative to suitably normalise bar formation times; we recommend this is done by the \emph{measured} orbital time $t_{\rm orb}$ at the disc scale radius $R_{\rm d}$.

Second, our live halos lack net rotation, restricting angular momentum transfer to a one-way flow from the disc to the halo. Previous work indicates that this exchange significantly affects both the onset and growth rate of stellar bars \citep[e.g.][]{Saha2013, Kataria2018, Collier2019, Collier2021}. Consequently, discs embedded in rotating halos may develop bars differently than those in non-rotating ones, even when other structural properties remain the same.

Finally, our galaxies evolve in perfect isolation, while real galaxies are subject to a cosmological environment including repeated mergers and tidal interactions throughout their history \citep[e.g.][]{Hammer2009, RodriguezGomez2015}. Numerous studies have demonstrated that such encounters can rapidly induce bar formation \citep[e.g.,][]{Noguchi1987, Berentzen2004, Lang2014, Lokas2014, Lokas2018}, due to the strong tidal forces acting on the disc \citep[see also][]{Mayer2004, RomanoDiaz2008, Martinez2017, Zana2018a, Peschken2019, RosasGuevara2024, Izquierdo2022, Frosst2025b}. We exclude these effects here to maintain controlled conditions and accurately quantify the intrinsic disc stability. In cosmological environments, where tidal perturbations and interactions are common, the bar formation times are possibly shorter and the parameter space for bar formation broader \citep[as found in][]{Frosst2025b}.

\subsection{Implications for high redshift bars}
Disc galaxies dominate the low-redshift Universe \citep[$z\lesssim1$; e.g.][]{Conselice2014}, but are also observed in abundance at much earlier cosmic times \citep[e.g.][]{Ferreira2023, EspejoSalcedo2025}. High redshift discs have been detected through ionised gas kinematics \citep[e.g.][]{Genzel2006}, Hubble Space Telescope (HST) observations \citep[e.g.][]{Johnson2017}, and increasingly with the James Webb Space Telescope \citep[JWST;][]{Guo2024}, in some cases out to $z\gtrsim 6$ \citep[e.g.][]{Kartaltepe2023}. Although many of these systems show ordered rotation, they are nevertheless typically thicker, more turbulent, and more frequently disturbed than their low-$z$ counterparts \citep{ForsterSchreiber2009,Stott2016,Birkin2024}. Remarkably, however, a non-negligible fraction of these high redshift discs host stellar bars \citep[][]{Guo2024, LeConte2024, EspejoSalcedo2025}, with such detections now extending to $z \approx 3-4$ \citep[e.g.][]{Huang2023, Costantin2023, Smail2023, Amvrosiadis2024}. 

Our results help explain the common occurrence of bars in high-redshift observations, which may seem surprising in light of the high turbulence (velocity dispersion) seen in many of these systems. Our simulations show that even discs with relatively high radial dispersions ($Q_{\rm min}=2.0$) can still form bars within a few Gyr provided they are sufficiently disc-dominated. With the growing observational evidence for many massive galaxies at higher redshifts being centrally baryon-dominated \citep[e.g.][]{Wuyts2016, Genzel2020, Price2021, Danhaive2025}, bars could be expected to be commonplace. Adding to this, external tidal perturbations caused by close encounters and mergers are more common at high redshift, which may further trigger bars in discs that would otherwise be stable against bar formation \citep[e.g.][]{Noguchi1987, RomanoDiaz2008, Lang2014, Zheng2025, Frosst2025b}.

Cosmological simulations provide a pathway to study bar growth at high redshifts \citep[e.g.][]{RosasGuevara2022, Lokas2025}, though their output cadence often lacks the time resolution needed to precisely measure $\tau_{\rm bar}$. Additionally, constant accretion and frequent interactions in these environments mean that bars may not follow a simple exponential growth phase \citep{Frosst2025b}. Despite these complexities, determining bar formation times in cosmological discs would enable a direct comparison to our idealised models, clarifying how $\tau_{\rm bar}$ scales with $f_{\rm disc}$ in realistic environments and identifying how disc kinematics might suppress bar formation.

\section{Summary}\label{sec:conclusions}
In this work, we examined the secular evolution of stellar bars using 145 self-consistent scale-free models of isolated galactic discs embedded in live and rigid dark matter haloes. These models were evolved using highly resolved dissipation free $N$-body simulations, following the methods of \citet{Frosst2024}, which offer converged bar properties with negligible numerical heating. We measured the bar growth as a function of time and quantified how variations in disc structure and kinematics affect bar formation. Our goal was to extend the analyses of \citetalias{Fujii2018} and \citetalias{BlandHawthorn2023} to also include thicker, more turbulent discs, more representative of those observed at higher redshift. Our main conclusions are as follows:
\begin{enumerate}
    \item All bars forming in our simulations grow exponentially until they experience buckling. Bar formation times are therefore best characterised by the exponential time-scale $\tau_{\rm bar}$. Unlike other standard metrics, such as the time where $A_2$ surpasses $0.2$, the exponential time-scale is independent of Poisson noise for sufficiently resolved systems.
    \item When normalised by the orbital time at the disc scale, $t_{\rm orb}$, the bar formation time-scale is well described by simple power-law relations with commonly used disc stability metrics, including the Ostriker–Peebles parameter $t_{\rm OP}$, the Efstathiou–Lake–Negroponte parameter $\epsilon_{\rm ELN}$, and the enclosed disc mass fraction $f_{\rm disc}$ (inside $2.2R_{\rm d}$). $f_{\rm disc}$ provides the most accurate predictor of $\tau_{\rm bar}/t_{\rm orb}$ (Eqn.~(\ref{eq:eta_fit}); Fig.~\ref{fig:original_ELN}, left). These power law relations are dimensionless and therefore universally applicable across a wide range of galaxy masses and sizes in models of isolated, collisionless stellar discs in co-evolving, non-rotating haloes. 
    \item Bars never form faster than $\tau_{\rm bar}\approx t_{\rm orb}$, but in other cases may form on timescales longer than a Hubble time.
    \item For galaxies with MW-like masses and sizes, stellar bars can form in less than a Hubble time provided $f_{\rm disc} \geq 0.18$, $t_{\rm OP} \geq 0.27$, and $\epsilon_{\rm ELN} \leq 1.44$ (Fig.~\ref{fig:original_ELN}).
    \item Significant velocity dispersion systematically slows down bar formation at fixed $f_{\rm disc}$ (Fig.~\ref{fig:QH_stability}). These effects can be incorporated into the power law relation through a simple empirical correction involving the dimensionless vertical dispersion $\tilde{\sigma}_{z}=\sigma_{z}/V_{\phi}$ (Eqn.~\ref{eq:correction}), yielding an practical, scale-free description of $\tau_{\rm bar}/t_{\rm orb}$ across all dispersion regimes explored. 
    \item All results above apply to galaxies in live haloes, which can co-evolve with the galaxies. If replaced by static halo potentials, stellar bars generally form more slowly (by about a factor~$2$) and, in some cases, may be suppressed altogether.
\end{enumerate}

By revisiting classical stability criteria and by introducing a new estimator that accounts for stellar velocity dispersions, this work extends existing studies of secular bar formation into regimes more representative of structurally diverse galaxies across cosmic times. While our analysis has been limited to idealised, isolated systems, it underscores the importance of considering both vertical and radial disc structure when characterising the assembly phase of bar formation. Future work should test these findings in full-physics simulations, including the effects of gas discs, central bulge components, and realistic star formation and feedback models. Furthermore, our results should be investigated in cosmological simulations where environmental effects and the broader assembly history of galaxies have been shown to shape the pathways to bar formation. In this way, the results presented here serve as a step toward a more general predictive theory of bar growth that can be connected directly with observations across cosmic time.

\section*{Acknowledgements}
DO acknowledge financial support from the Australian Research Council through their Future Fellowship scheme (project number FT190100083). This work was supported by resources provided by the Pawsey Supercomputing Research Centre’s Setonix Supercomputer (\url{https://doi.org/10.48569/18sb-8s43}), with funding from the Australian Government and the Government of Western Australia. This research was undertaken with the assistance of resources and services from the National Computational Infrastructure (NCI), which is supported by the Australian Government. This work has benefited from the following public \textsc{Python} packages: \textsc{AGAMA} \citep{Vasiliev2019} \textsc{Scipy} \citep{Virtanen2020}, \textsc{Numpy} \citep{Harris2020}, and \textsc{Matplotlib} \citep{Hunter2007}.

\section*{Data Availability}
Our data is available upon reasonable request, or is otherwise obtainable from publicly available codes. 
 

\bibliographystyle{mnras}
\bibliography{references} 



\appendix

\section{Bar Strength Evolution}
\begin{figure}
    \includegraphics[width=\columnwidth]{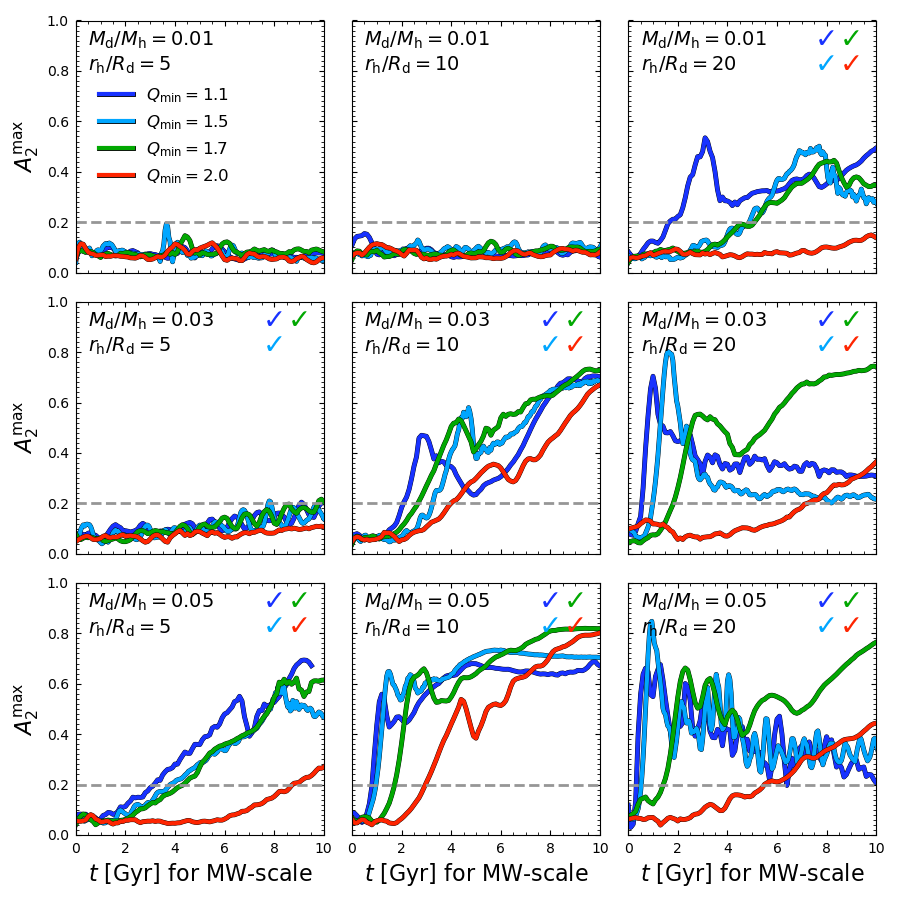}
    \caption{
        The bar strength, $A_{2}^{\rm max}$ measured at each snapshot, plotted as a function of time for model spanning $M_{\rm d}/M_{\rm h} \in \{0.01, 0.03, 0.05\}$ and $C \in \{5,10,20\}$. 
        The profiles are coloured based on the initial value of $Q_{\rm min}$ following Fig.~\ref{fig:QH_stability}; check-marks of the same colour indicate whether bar formation occurs within the simulation runtime. The horizontal grey dashed line displays the delineation between barred and unbarred systems. 
    }
    \label{fig:varyQall}
\end{figure}

\begin{figure}
    \includegraphics[width=\columnwidth]{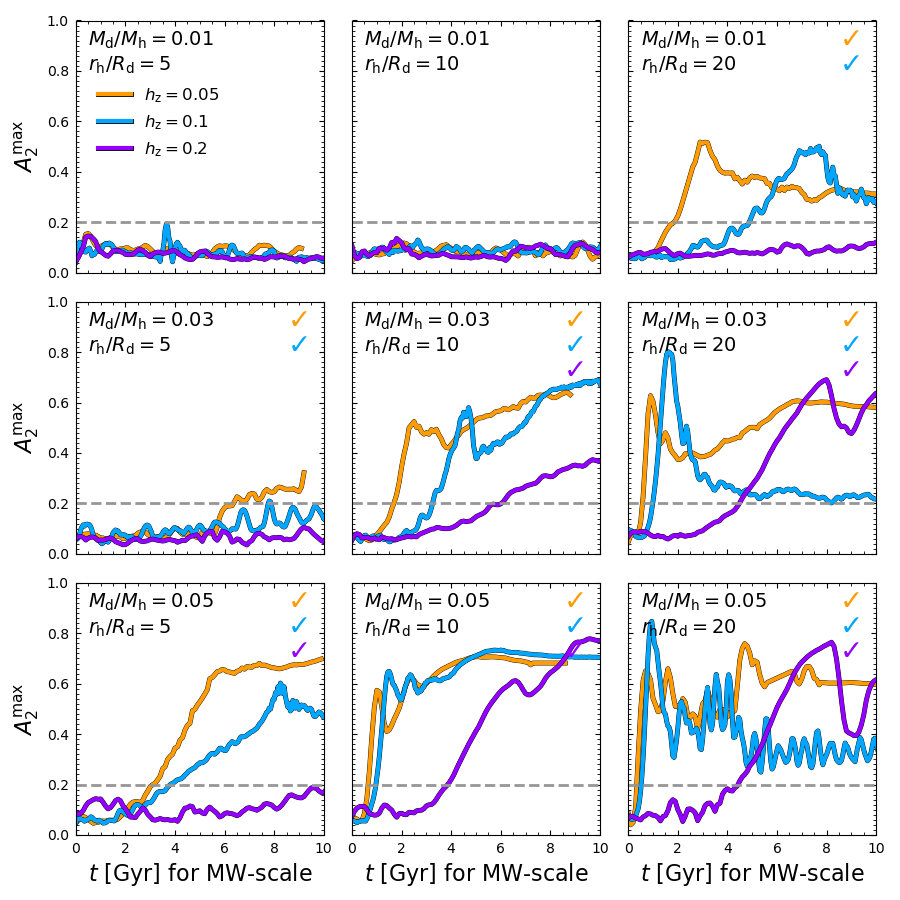}
    \caption{
        The bar strength, $A_{2}^{\rm max}$ measured at each snapshot, , plotted as a function of time for model spanning $M_{\rm d}/M_{\rm h} \in \{0.01, 0.03, 0.05\}$ and $C \in \{5,10,20\}$. 
        The profiles are coloured based on the initial value of $h_{z}$ following Fig.~\ref{fig:QH_stability}; check-marks of the same colour indicate whether bar formation occurs within the simulation runtime. The horizontal grey dashed line displays the delineation between barred and unbarred systems. 
    }
    \label{fig:varyHzall}
\end{figure}

In Fig.~\ref{fig:varyQall} and Fig.~\ref{fig:varyHzall} we present the time evolution of $A_{2}^{\rm max}$ for the models with varied $Q_{\rm min}$ and models with varied $h_{z}$, respectively. Each bar formation history is subsequently fit with Eqn.~\ref{eq:taubar}. These metrics are then used to quantitatively inform us of the stability of these models to bar formation. Bar formation in individual mass models is delayed as the discs become more radially dispersed (i.e., as $Q_{\rm min}$ increases), and as it becomes more vertically dispersed (i.e., as $h_{z}$ increases). Precise descriptions of these effects are available in the text. 

\bsp	
\label{lastpage}
\end{document}